\documentclass[a4paper,11pt]{article}

\usepackage{contribution}

\newcommand{\weblink}[2][]{%
    \ifthenelse{\equal{#1}{}}%
    {\textnormal{\url{#2}}}%
    {\textnormal{\href{#2}{#1}}}%
}

\def\beq{\begin{equation}}
\def\eeq#1{\label{#1}\end{equation}}
\def\eeqn{\end{equation}}

\def\beqa{\begin{eqnarray}}
\def\eeqa#1{\label{#1}\end{eqnarray}}
\def\eeqan{\end{eqnarray}}

\let\bar=\overbar

\def\Dslash{\not{\hbox{\kern-4pt $D$}}}
\def\dslash{\not{\hbox{\kern-2pt $\del$}}}

\def\msb{{\bar{\ssstyle M \kern -1pt S}}}

\usepackage{bm}

\newcommand{\contribution}[7][]{%
  \clearpage
  \thispagestyle{plain}
  \ifthenelse{\equal{#1}{}}
  {\hypersetup{pdftitle={#2}}}
  {\hypersetup{pdftitle={#1}}}
  \hypersetup{pdfauthor={{#3} {#4}}}
  {\centering\normalfont\LARGE\bfseries\sffamily #2 \par\nobreak}
  \lhead{}
  \chead{%
    \textit{\footnotesize XIV International Conference on Hadron Spectroscopy
      (\weblink[\textit{hadron2011}]{http://www.hadron2011.de}), 13-17 June 2011, Munich, Germany}%
  }
  \rhead{}
  \bigskip
  \begin{center}
    {#3} {#4}\ifthenelse{\equal{#6}{}}{}{\footnote{\weblink[#6]{mailto:#6}}}
    \ifthenelse{\equal{#7}{}}{}{#7} \\
    \textit{#5}
  \end{center}
  \bigskip
}

\renewcommand{\abstract}[1]{%
  \begin{center}
    \begin{minipage}{0.85\textwidth}
      \begin{footnotesize}
        #1
      \end{footnotesize}
    \end{minipage}
  \end{center}
  \bigskip
}

\begin{document}

%
%
%
%
%
{  



%

\contribution[]  
{HEAVY QUARKONIA\\Recent Results from CLEO}  
{Kamal K.}{Seth}  
{Northwestern University \\
  Evanston, IL, USA }  
{kseth@northwestern.edu}  
{}  
%

%

\section{Introduction}

Before it stopped data taking in 2008, CLEO had accumulated a large amount of $e^+e^-$ data in the bottomonium and charmonium regions, as shown in Table~\ref{t1}. These data have led to valuable contributions in the spectroscopy of both $|c\bar{c}\rangle$ and $|b\bar{b}\rangle$ quarkonia, and their continuing analysis is leading to new physics results. In this presentation I want to describe some of the results obtained since HADRON 2009~\cite{c1}. More than a dozen papers on spectroscopy have been published since then, and my choice for this time-limited presentation is necessarily a subjective one.

\begin{table}[h]
\begin{center}
\begin{tabular}{|l|l|}
\hline \hline
\textbf{Charmonium region}  &  \textbf{Bottomonium region} \\ \hline\hline
$\psi(2S,3686):54~\mathrm{pb^{-1}},\sim 27~\mathrm{million}~\psi(2S)$ & $\Upsilon(1S):1056~\mathrm{pb^{-1}},20.8~\mathrm{million}~\Upsilon(1S)$ \\\hline
$\psi(3770):818~\mathrm{pb^{-1}},\sim 5~\mathrm{million}~\psi(3770)$  & $\Upsilon(2S):1305~\mathrm{pb^{-1}},9.3~\mathrm{million}~\Upsilon(2S)$ \\\hline
$\psi(4170):586~\mathrm{pb^{-1}},\sim 5~\mathrm{million}~\psi(4170)$  & $\Upsilon(3S):1378~\mathrm{pb^{-1}},5.9~\mathrm{million}~\Upsilon(3S)$ \\\hline
$\sqrt{s}=3670~\mathrm{MeV}:21~\mathrm{pb^{-1}}$                      & $\Upsilon(4S):9400~\mathrm{pb^{-1}},15.4~\mathrm{million}~B\bar{B}$ \\\hline
$\sqrt{s}=4040~\mathrm{MeV}:20.7~\mathrm{pb^{-1}}$                    & $\sqrt{s}=10,520~\mathrm{MeV}:4500~\mathrm{pb^{-1}}$ \\\hline
$\sqrt{s}=4260~\mathrm{MeV}:13.2~\mathrm{pb^{-1}}$                    & Off $\Upsilon(nS):800~\mathrm{pb^{-1}}$ \\\hline\hline
\end{tabular}
\end{center}
\caption{CLEO data in the charmonium and bottomonium regions}
\label{t1}
\end{table}

\section{Hyperfine Interaction in Quarkonia}

One of our major interests at CLEO during the last five years has been in the study of the hyperfine interaction in quarkonia, and our investigations into it have continued to yield new insights into the subject.\\

The hyperfine or spin-spin interaction in $|c\bar{c}\rangle$ and $|b\bar{b}\rangle$ quarkonia leads to the hyperfine splitting between spin-triplet and spin-singlet states, which is defined as
$$\Delta M_{hf}(nL)=M(n^3L)-M(n^1L)$$
where n and L are the principal and angular momentum quantum numbers. \\

For a purely Coulombic central potential, as for $|e^+e^-\rangle$ positronium, for $|q\bar{q}\rangle$ quarkonium the hyperfine interaction is a contact interaction, and leads to the predictions ---
$$\Delta M_{hf}(nS)=M(n^3S_1)-M(n^1S_0)=\frac{32\pi\alpha_s(m_q)}{9}(\psi(0)/m_q)^2,\quad L=0$$
$$\Delta M_{hf}(nL)=M(n^3L)-M(n^1L)=0, \quad\qquad\qquad\qquad\qquad\qquad L\neq0$$
where $\alpha_s(m_q)$ is the strong coupling constant for quark mass $m_q$, and $\psi(0)$ is the wave function at the origin.\\

The interest for quarkonia is in determining the extent to which these predictions are valid, because for quarkonia the central potential has the confinement part in addition to the Coulombic part, and the charm and beauty quarks, which have different masses, bring in different relativistic and higher order effects. \\

As is well known, in $e^+e^-$ annihilation the spin-triplet-S wave states, called $\psi_c$ and $\Upsilon_b$, $^3S_1(J^{PC}=1^{--})$ are directly produced, and the spin-triplet P-wave states, $^3P_J(J^{PC}=0^{++},1^{++},2^{++})$, called $\chi_{cJ}$,$\chi_{bJ}$, are strongly excited by E1 radiative transitions from the triplet S states. In contrast, the M1 radiative transitions to the spin-singlet states, $^1S_0(J^{PC}=0^{-+})$, called $\eta_c$ and $\eta_b$, and $^1P_1(J^{PC}=1^{+-})$, called $h_c$ and $h_b$, are much weaker and much more difficult to identify because of their close proximity to the triplet states. As a result, for more than two decades after the discovery of $J/\psi(1^3S_1)_{c\bar{c}}$, $\psi(2^3S_1)_{c\bar{c}}$, and $\Upsilon(n^3S_1)_{b\bar{b}}(n=1,2,3,4)$, the only singlet state which was successfully identified was $\eta_c(1^1S_0)$, and the only hyperfine splitting which was known was $\Delta M_{hf}(1S)_{c\bar{c}}=116.6\pm1.2~\mathrm{MeV}$~\cite{c2}. As a result, it was not known how the hyperfine interaction between quarks changes with greater exposure to the confinement potential with increasing redius (1S versus 2S), with increasing angular momentum (S-wave versus P-wave), and increasing quark mass (c-quarks versus b-quarks). Great progress in answering these questions has been recently made by B-factories and CLEO in challenging new measurements. Belle identified $\eta_c'(2^1S_0)$ in B-decays~\cite{c3}, and it was confirmed by CLEO~\cite{c4} and BaBar~\cite{c5} in two photon formation. CLEO~\cite{c6} identified $h_c(1^1P_1)$ in $\psi(2S)$ decay. BaBar~\cite{c7} identified $\eta_b(1^1S_0)$ in $\Upsilon(3S)$ decays, and it was confirmed by CLEO~\cite{c8}. Identification of $\eta_c(3^1S_0)$ and $h_c(2^1P_1)$, which lie above the $D\bar{D}$ break-up threshold, and $\eta_b(2,3^1S_0)$ and $h_b(1,2^1P_1)$ remained as challenges\footnote{For the breaking news on the discovery of $h_b(1,2^1P_1)$ see Ref~\cite{c14}.}.

\section{New Results for P-wave Singlet State $h_c(1^1P_1)$}

CLEO reported the discovery of $h_c$ in 2005~\cite{c6}, and the precision measurement of its mass in 2008~\cite{c9},
$$\mathrm{CLEO ~[2008]}:~M(h_c,1^1P_1)=3525.28\pm0.19\mathrm{(stat)}\pm0.12\mathrm{(syst)~MeV}.$$
It is extremely gratifying that BES III~\cite{c10} has now confirmed this, with the result:
$$\mathrm{BES~III ~[2010]}:~M(h_c,1^1P_1)=3525.40\pm0.13\mathrm{(stat)}\pm0.18\mathrm{(syst)~MeV}.$$
The centroid of the $^3P_J$ states $(\chi_{0,1,2})$ is known to be \cite{c2}
$$\langle M(^3P_J)\rangle=[5M(^3P_2)+3M(^3P_1)+M(^3P_0)]=3525.30\pm0.04~\mathrm{MeV}.$$

If the $^3P_J$ states centroid mass $\langle M(^3P_J)\rangle$ above is identified as the mass $M(^3P)$, then the hyperfine splittings are 
\begin{center}
CLEO:$\quad\Delta M_{hf}(1P)_{c\bar{c}}=+0.02\pm0.23~\mathrm{MeV},$ and 
\end{center}
\begin{center}
BES III:$\quad\Delta M_{hf}(1P)_{c\bar{c}}=-0.10\pm0.22~\mathrm{MeV}.\qquad$
\end{center}
However, it must be pointed out that the identification of the centroid is only valid if the spin-orbit splitting between the $^3P_J$ states in perturbatively small. This is hardly the case here with $M(^3P_2)-M(^3P_0)=141.45\pm0.32~\mathrm{MeV}$, and the perturbative prediction $M(^3P_1)-M(^3P_0)=(5/2)\times[M(^3P_2)-M(^3P_1)]=114~\mathrm{MeV}$ is $20\%$ larger than the experimental result = 96 MeV. Why then is $\Delta M_{hf}(1P)$ so very close to zero? It is a mystery.

\subsection{Beyond the discovery of $h_c$~\cite{c11}}

\begin{figure}
\begin{center}
\includegraphics[width=2.3in]{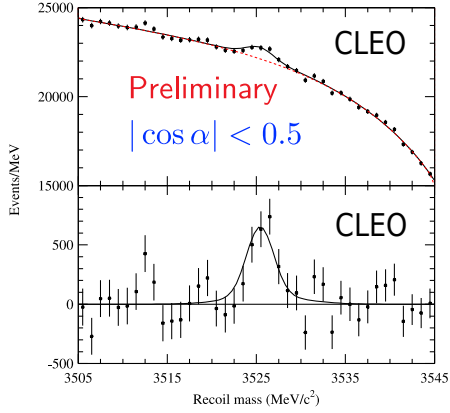}
\caption{New CLEO result for the inclusive analysis of $\psi(2S)\rightarrow\pi^0 h_c$}
\end{center}
\end{figure}

In our $h_c$ discovery and mass papers in the decay
$$\psi(2S)\rightarrow\pi^0 h_c,~h_c\rightarrow\gamma\eta_c$$
we made inclusive analyses of the $\pi^0$ recoil spectrum by either constraining the $\gamma$ energy or $\eta_c$ mass. As a result we could only determine the product branching fraction $\mathcal{B}(\psi(2S)\rightarrow\pi^0h_c)\times\mathcal{B}(h_c\rightarrow\gamma\eta_c)$.\\

BES III data for 100 million $\psi(2S)$ allowed them to observe $h_c$ directly in the $\pi^0$ recoil spectrum. It occured to us at CLEO recently to also attempt to also identify $h_c$ directly in the $\pi^0$ recoil spectrum despite our factor four smaller 25.9 million $\psi(2S)$ sample. By rejecting very asymmetric $\pi^0\rightarrow2\gamma$ decays, as shown in Fig.~1, we were successful in identifying $h_c$. Our result is in excellent agreement with the BES III result~\cite{c10}

\begin{tabbing}
~~~~~~\=CLEO:\quad $\mathcal{B}[\psi(2S)\rightarrow\pi^0h_c]=(9.0\pm1.5\pm1.2)\times10^{-4}$~\cite{c11} \\
~~~\\
\>BES III:\quad $\mathcal{B}[\psi(2S)\rightarrow\pi^0h_c]=(8.4\pm1.3\pm1.0)\times10^{-4}$~\cite{c10} \\
~~~\\
\>Average:\quad $\mathcal{B}[\psi(2S)\rightarrow\pi^0h_c]=(8.7\pm1.2)\times10^{-4}$ 
\end{tabbing}

\subsection{Hadronic decays of $h_c$~\cite{c12}}

The CLEO result~\cite{c9} $\mathcal{B}_1(\psi(2S)\rightarrow\pi^0h_c)\times\mathcal{B}_2(h_c\rightarrow\gamma\eta_c)=(4.19\pm0.55)\times10^{-4}$ has also been confirmed by BES III~\cite{c10} with $\mathcal{B}_1(\psi(2S)\rightarrow\pi^0h_c)\times\mathcal{B}_2(h_c\rightarrow\gamma\eta_c)=(4.58\pm0.64)\times10^{-4}$, and the average is $\mathcal{B}_1(\psi(2S)\rightarrow\pi^0h_c)\times\mathcal{B}_2(h_c\rightarrow\gamma\eta_c)=(4.39\pm0.42)\times10^{-4}$. Combined with $\mathcal{B}_1(\psi(2S)\rightarrow\pi^0h_c)=(8.7\pm1.2)\times10^{-4}$, we obtain ${B}_2(h_c\rightarrow\gamma\eta_c)=(50.5\pm8.5)\%$. Therefore, we expect that the remaining $50\%$ decays of $h_c$ must be to hadrons. This suggests that decays to odd number of pions may be an important component of the hadronic decays. We have therefore measured~\cite{c12}
$$\psi(2S)\rightarrow\pi^0 h_c,~h_c\rightarrow(\pi^+\pi^-)\pi^0,~n=1,2,3$$
Unfortunately, no significant yield was found for 3 or 7 pion final states. Only a small 5 pion transition was observed with
$$\mathcal{B}(h_c\rightarrow2(\pi^+\pi^-)\pi^0)=(1.9^{+0.7}_{-0.5})\times10^{-5}$$
This leaves us with the interesting question of what are the $\sim50\%$ unobserved hadronic decays of $h_c$.

\subsection{Discovery of a new mode of $h_c$ production~\cite{c13}}

\begin{figure}
\begin{center}
\includegraphics[width=2.3in]{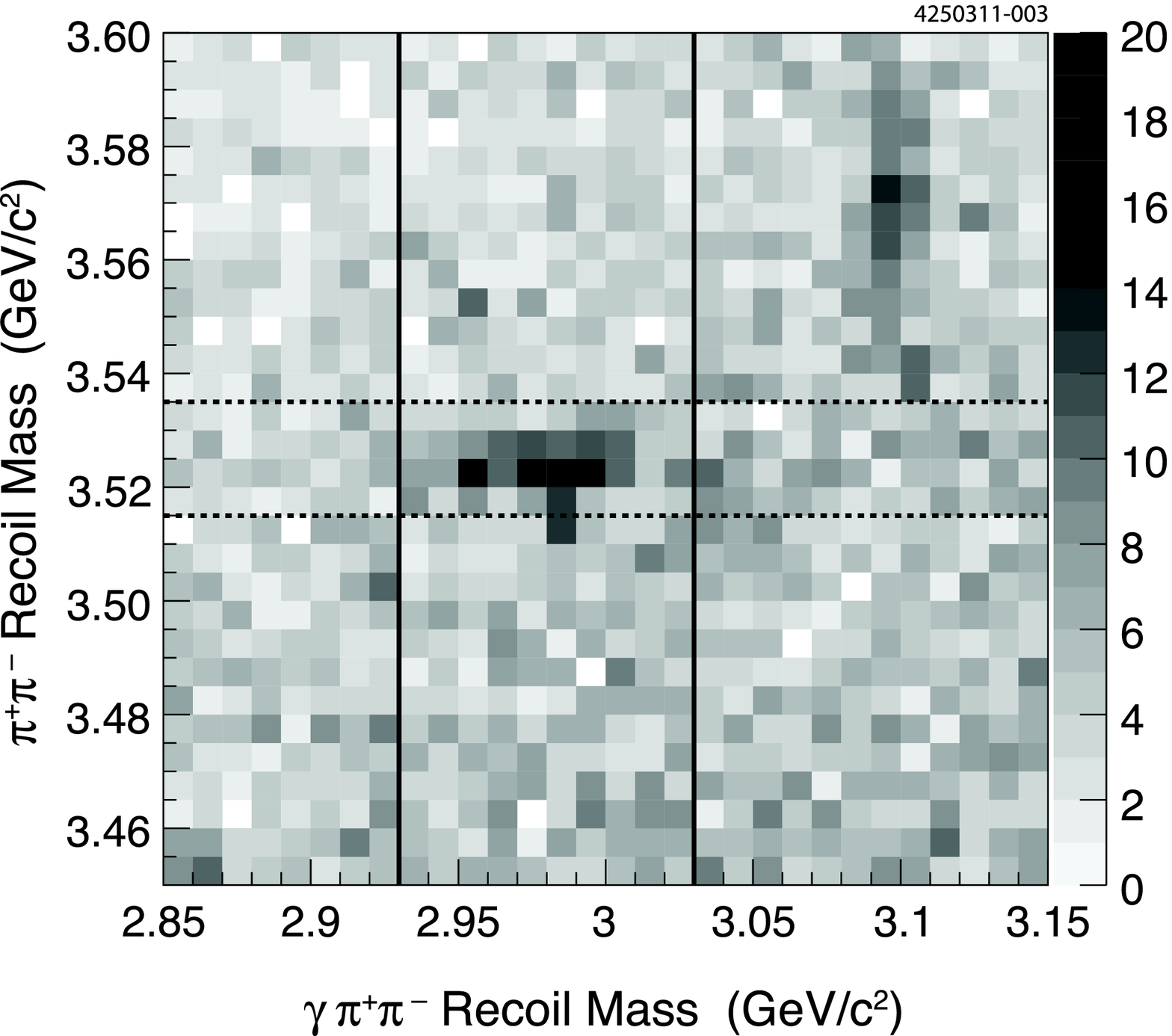}
\qquad
\includegraphics[width=2.5in]{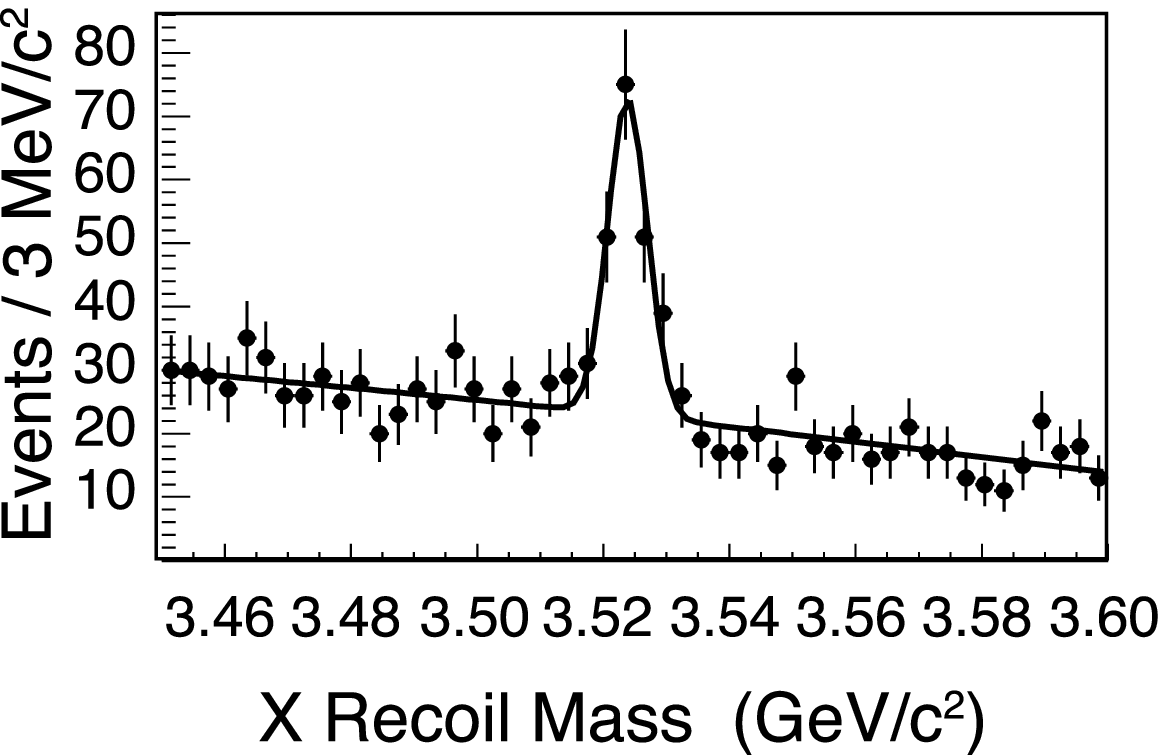}
\end{center}
\caption{Illustrating the identification of $h_c$ in $e^+e^-(4170)\rightarrow\pi^+\pi^-h_c$. Left: two dimensional plot showing $h_c$ enhancement at the intersection of $M(\eta_c)\approx2.98~\mathrm{GeV}$ and $M(h_c)\approx3.52~\mathrm{GeV}$. Right: distribution of events in the box marked in the two dimensional plot as function of $\pi^+\pi^-$ recoil mass.}
\end{figure}

CLEO has made an important discovery in identifying $h_c$ formation in the $\pi^+\pi^-$ decay of $\psi(4170)$ above the $D\bar{D}$ threshold~\cite{c13}. Using $586~\mathrm{pb^{-1}}$ of $e^+e^-$ annihilation data at $\sqrt{s}=4170~\mathrm{MeV}$ we observe a $10\sigma$ signal for $h_c$ in the decay
\begin{center}
$e^+e^-(4170)\rightarrow\pi^+\pi^-h_c(1P),$
\end{center}
with $h_c\rightarrow\gamma\eta_c, \eta_c\rightarrow$ 12 decay modes\footnote{$\eta_c\rightarrow2(\pi^+\pi^-),2(\pi^+\pi^-)2\pi^0,3(\pi^+\pi^-),K^{\pm}K^0_S\pi^{\mp},K^{\pm}K^0_S\pi^{\mp}\pi^+\pi^-,K^+K^-\pi^0,K^+K^-\pi^+\pi^-,K^+K^-\pi^+\pi^-\pi^0,$\\
$K^+K^-2(\pi^+\pi^-),2(K^+K^-),\eta\pi^+\pi^-,$ and $\eta2(\pi^+\pi^-)$.}.\\

In the two dimensional plot shown in Fig. 2 the $h_c$ signal is clearly seen in $\pi^+\pi^-$ recoil mass at the intersection of its radiative decay to $\eta_c$ at 2.98 GeV. (The enhancement at 3.1 GeV is due to $J/\psi$.) In the projection $h_c$ is seen as a strong enhancement over a featureless background. The production cross section is a very healthy $15.6\pm4.2~\mathrm{pb}$. The paper has been accepted for publication in the PRL.\\

Our discovery of the population of $h_c(1P)$ in $e^+e^-$ annihilations above the $D\bar{D}$ threshold of charmonium has led the Belle collaboration to search for $h_b(1P,2P)$ in $e^+e^-$ annihilations at $\sqrt{s}=10.685~\mathrm{GeV}$ using the same technique of recoil against $\pi^+\pi^-$. They have achieved dramatic success, as you have already heard in their plenary presentation~\cite{c14}.

\section{Hadronic Decays of P-wave States of Bottomonium~\cite{c16}}

Compared to charmonium very few decays of bottomonium states have ever been measured. Earlier we reported on the first measurements of $\chi_{bJ}(1P,2P),J=0,1,2,$ decays to fourteen exlusive hadronic final states~\cite{c15}.
$$\Upsilon(2S,3S)\rightarrow\gamma\chi_{bJ}(1P,2P),\quad\chi_{bJ}(1P,2P)\rightarrow\mathrm{hadrons}$$
We have now made the first measurements of
$$\Upsilon(2S,3S)\rightarrow\gamma\chi_{bJ}(1P)\rightarrow\gamma\gamma\Upsilon(1S)~[16]$$
The results from $\Upsilon(2S)\rightarrow\gamma\chi_{bJ}(1P)$ are
$$\mathcal{B}[\chi_{bJ}(1P)\rightarrow\gamma\Upsilon(1S)]~\mathrm{in}~\%=1.73\pm0.35(\chi_0),~~ 33.0\pm2.6(\chi_1),~~ 18.5\pm1.4(\chi_2)$$
These measurements lead to much improved determinations of
\begin{center}
$\mathcal{B}[\Upsilon(3S)\rightarrow\gamma\chi_{b1}(1P)]=(1.63\pm0.46)\times10^{-3}$ (CLEO), $\quad<1.9\times10^{-3}$~[2,PDG]
\end{center}
\begin{center}
$\mathcal{B}[\Upsilon(3S)\rightarrow\gamma\chi_{b2}(1P)]=(7.7\pm1.3)\times10^{-3}$ (CLEO), $\quad<20.3\times10^{-3}$~[2,PDG]
\end{center}

\section{Decays of $\psi(2S)$ to $p\bar{p}+\gamma,\pi^0$ and $\eta$, and search for baryonium in $\psi(2S)$ and $J/\psi$ decays~\cite{c17}}

\begin{figure}
\begin{center}
\includegraphics[width=5.6in]{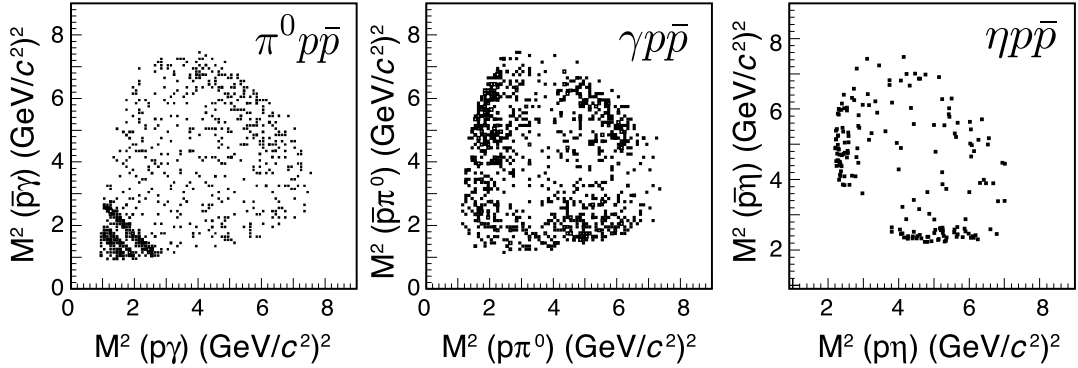}
\end{center}
\caption{Dalitz plots for the three decays, $\psi(2S)\rightarrow\pi^0p\bar{p},\gamma p\bar{p}$, and $\eta p\bar{p}$.}
\end{figure}

This CLEO investigation~\cite{c17} was motivated by the longstanding claim by BES for the interpretation of an observed near-threshold enhancement in the decay, $J/\psi\rightarrow\gamma(p\bar{p})$ as evidence for a weakly bound proton-antiproton resonance, $\mathrm{R_{thr}}$, with $M(p\bar{p})=1859^{+6}_{-27}~\mathrm{MeV}$, $\Gamma<30~\mathrm{MeV}$, and 
$$\mathcal{B}(J/\psi\rightarrow\gamma\mathrm{R_{thr}})\times\mathcal{B}(\mathrm{R_{thr}}\rightarrow p\bar{p})=(7.0^{+1.9}_{-0.9})\times10^{-5}.$$

We argued that if the baryonium resonance was real, it should also be seen in $\psi(2S)\rightarrow\gamma(p\bar{p})$, and perhaps also in $\pi^0(p\bar{p})$ and $\eta(p\bar{p})$. Accordingly, we made a detailed analysis of our data set of 24.5 million $\psi(2S)$. The Dalitz plots in Fig. 3 show that a number of light quark resonances are excited in all three decays.\\

The structures observed in the Dalitz plots were analyzed via their projections as shown in Fig. 4. As listed in Table~\ref{t2}, branching fractions were determined for a number of baryon ($N^*$), and meson resonances ($R$) which decay into $p\bar{p}$. Most of these represent first such measurements. We note that among the intermediate states identified are $f_2(2150)$ and $N^*(2300)$ which have since been also observed by BES III [Hai-Bo Li at this conference]. \\

\begin{figure}
\begin{center}
\includegraphics[width=1.8in]{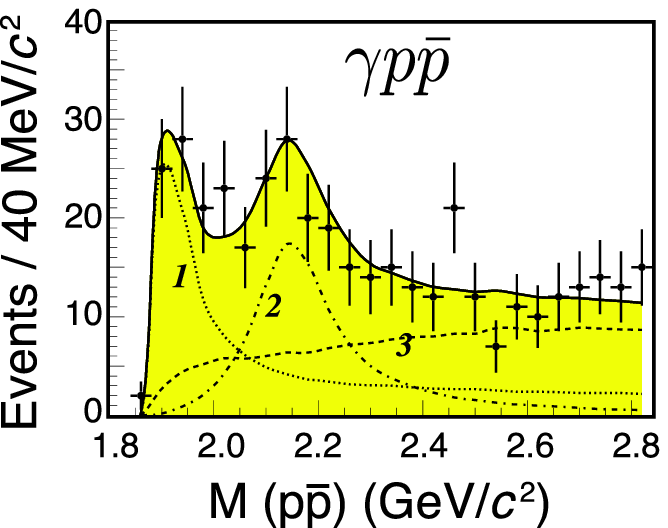}
~~~~
\includegraphics[width=1.8in]{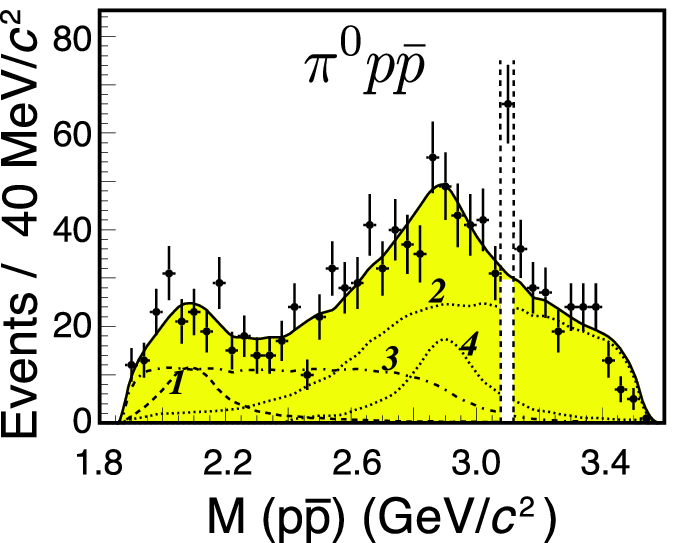}
~~~~
\includegraphics[width=1.8in]{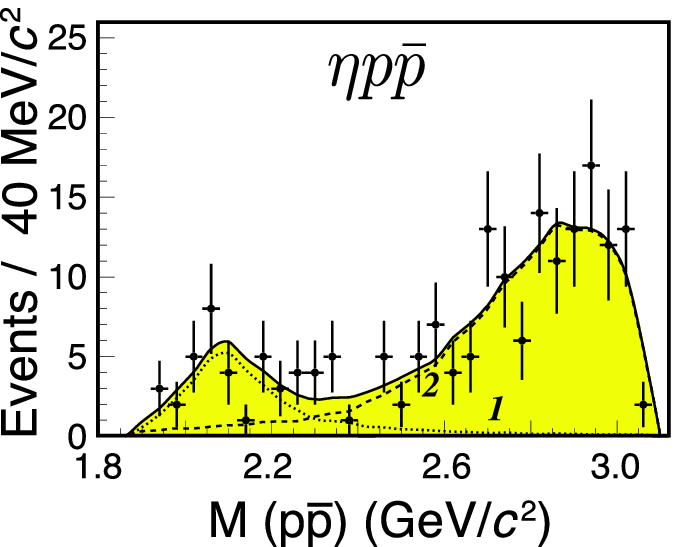}
\end{center}
\caption{Event projections as function of $M(p\bar{p})$ in the decays $\psi(2S)\rightarrow\pi^0p\bar{p},\gamma p\bar{p}$, and $\eta p\bar{p}$. The curves show contributions of different intermediate resonances and their total.}
\end{figure}

\begin{table}[!tb]
\begin{center}
\begin{tabular}{|l|c|c|} \hline\hline
Quantity  &   CLEO $(10^{-5})$  &   PDG10 $(10^{-5})$ \\ \hline
$\mathcal{B}(\psi(2S)\rightarrow\gamma p\bar{p})$        &    $4.18\pm0.3$     &      $2.9\pm0.6$    \\
$\mathcal{B}(\psi(2S)\rightarrow\pi^0 p\bar{p})$        &    $15.4\pm0.9$     &      $13.3\pm1.7$    \\
$\mathcal{B}(\psi(2S)\rightarrow\eta p\bar{p})$        &    $5.6\pm0.7$     &      $6.0\pm1.2$    \\

$\mathcal{B}(\psi(2S)\rightarrow\gamma f_2(1950))\times\mathcal{B}(f_2(1950)\rightarrow p\bar{p})$        &    $1.2\pm0.2$     &      \\
$\mathcal{B}(\psi(2S)\rightarrow\gamma f_2(2150))\times\mathcal{B}(f_2(2150)\rightarrow p\bar{p})$        &    $0.72\pm0.18$     &      \\

$\mathcal{B}(\psi(2S)\rightarrow\pi^0 R_1(2100))\times\mathcal{B}(R_1(2100)\rightarrow p\bar{p})$        &    $1.1\pm0.4$     &      \\
$\mathcal{B}(\psi(2S)\rightarrow\pi^0 R_2(2900))\times\mathcal{B}(R_2(2900)\rightarrow p\bar{p})$        &    $2.3\pm0.7$     &      \\
$\mathcal{B}(\psi(2S)\rightarrow\eta R_1(2100))\times\mathcal{B}(R_1(2100)\rightarrow p\bar{p})$        &    $1.2\pm0.4$     &      \\
$\mathcal{B}(\psi(2S)\rightarrow\bar{p} N_1^*(1440))\times\mathcal{B}(N_1^*(1440)\rightarrow p\pi^0)$        &    $8.1\pm0.8$     &      \\
$\mathcal{B}(\psi(2S)\rightarrow\bar{p} N_2^*(2300))\times\mathcal{B}(N_2^*(2300)\rightarrow p\pi^0)$        &    $4.0\pm0.6$     &      \\
$\mathcal{B}(\psi(2S)\rightarrow\bar{p} N^*(1535))\times\mathcal{B}(N^*(1535)\rightarrow p\eta)$        &    $4.4\pm0.7$     &      \\
\hline\hline
\end{tabular}
\end{center}
\caption{Branching fractions determined for $\psi(2S)$ decays into various intermediate $N^*$ and meson states $R_n$ which decay to $p\bar{p}$.}
\label{t2}
\end{table}

We now turn to our results for the search for $p\bar{p}$ threshold enhancements. These are illustrated in Fig. 5 in terms of $\Delta M=M(p\bar{p})-2m_p$.\\

$\bm{\psi(2S)\rightarrow\gamma p\bar{p}}$: As shown in Fig. 5, we find no evidence for a threshold enhancement in $M(p\bar{p})$, and establish the upper limit for a resonance with BES parameters for $R_{\mathrm{thr}}$
$$\mathcal{B}(\psi(2S)\rightarrow\gamma\mathrm{R_{thr}})\times\mathcal{B}(\mathrm{R_{thr}}\rightarrow p\bar{p})<1.6\times10^{-6}.$$

$\bm{J/\psi\rightarrow\gamma p\bar{p}}$: Using the data for 8.7 million $J/\psi$ produced via $\psi(2S)\rightarrow\pi^+\pi^-J/\psi$, $\mathrm{R_{thr}}$ was also searched for in $J/\psi\rightarrow\gamma (p\bar{p})$. \\

As shown in Fig.~5, the observed threshold enhancement fitted in the region, $\Delta M=0-900~\mathrm{MeV}$ leads to
\begin{center}
$M(\mathrm{R_{thr}})=1837\pm14~\mathrm{MeV},\Gamma(\mathrm{R_{thr}})=0^{+44}_{-0}~\mathrm{MeV}$, and
\end{center}

\begin{center}
$\mathcal{B}(J/\psi\rightarrow\gamma\mathrm{R_{thr}})\times\mathcal{B}(\mathrm{R_{thr}}\rightarrow p\bar{p})=(11.4^{+6.0}_{-4.0})\times10^{-5}.$
\end{center}

\begin{figure}
\begin{center}
\includegraphics[width=2.3in]{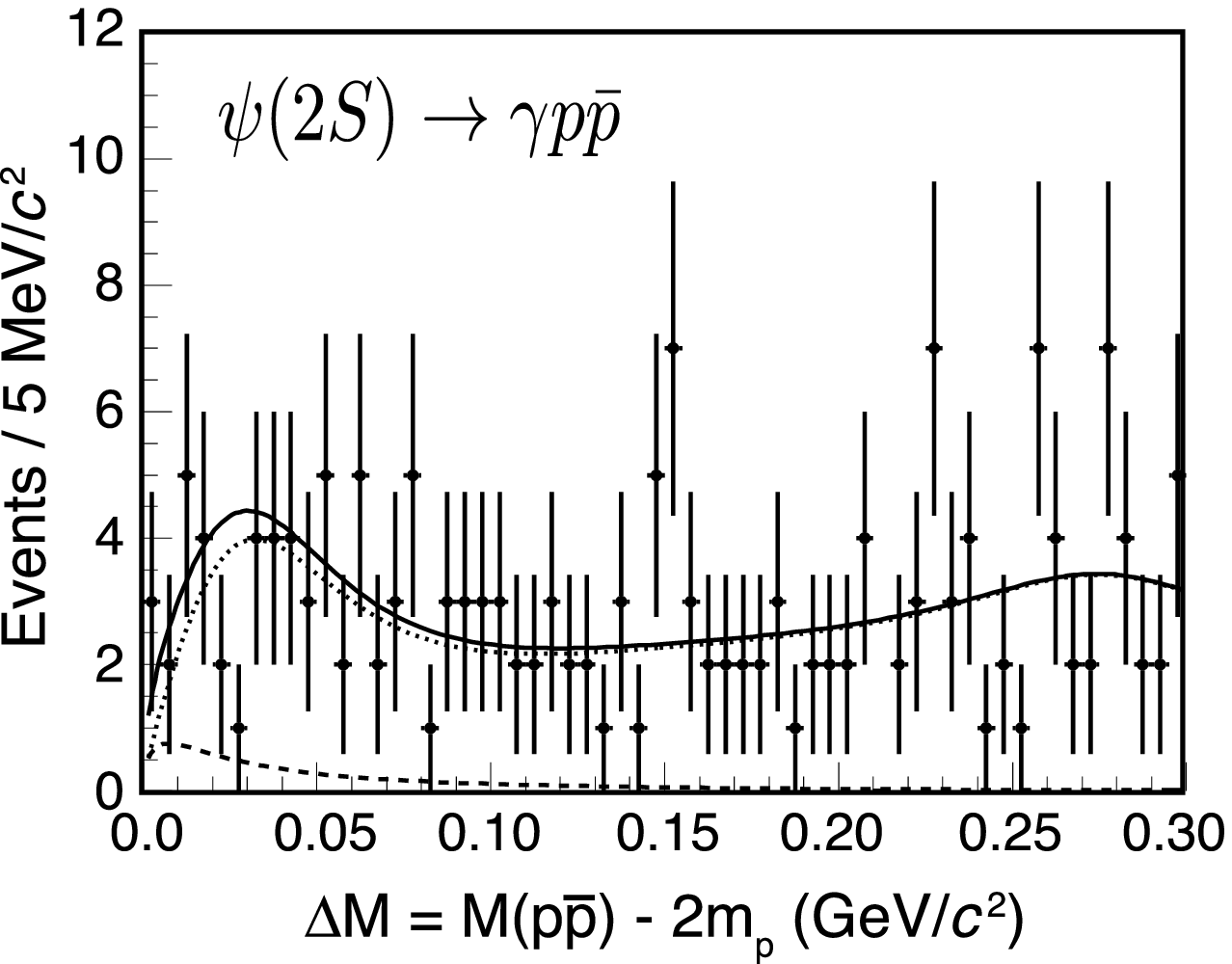}
\quad
\includegraphics[width=2.3in]{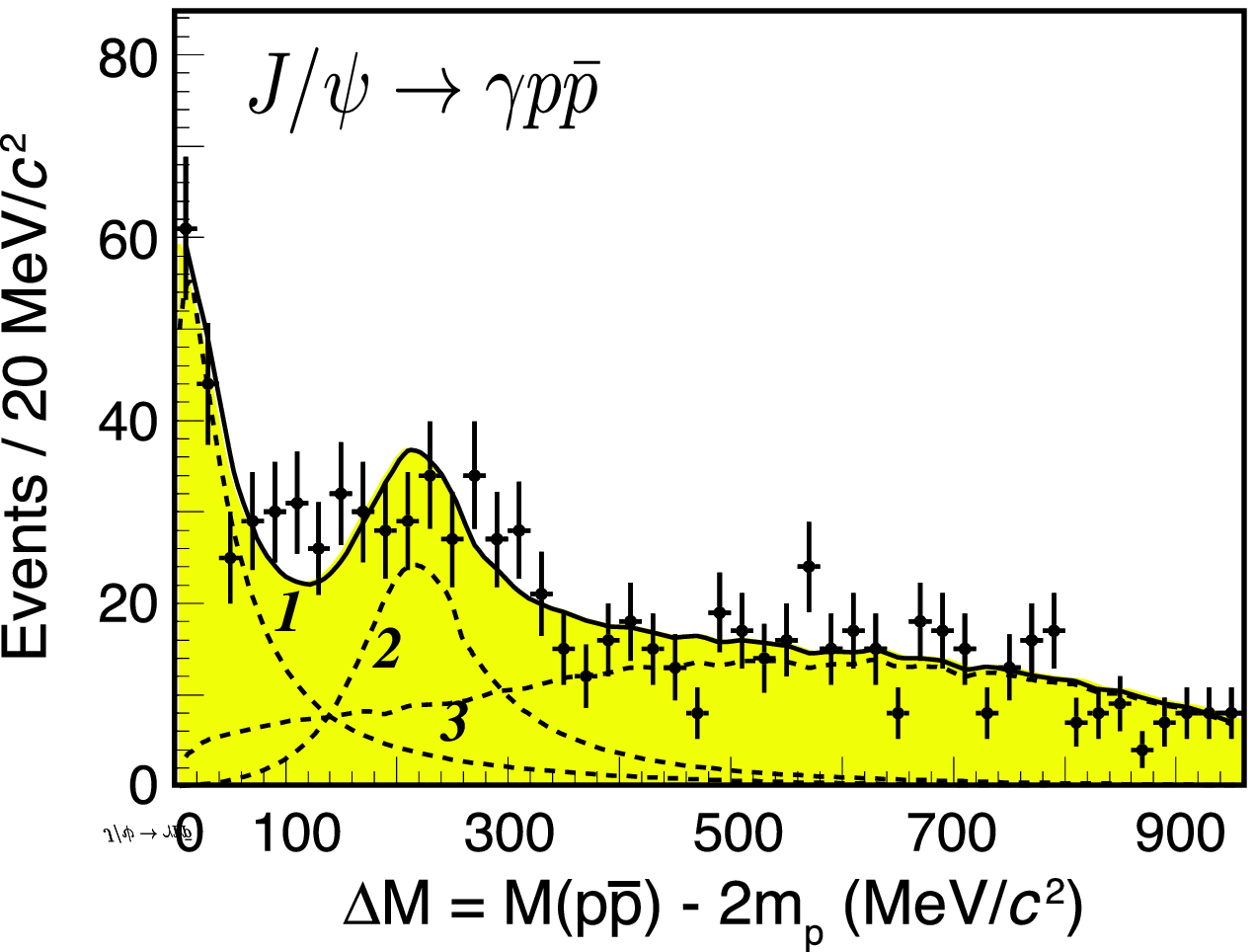}
\caption{Event distributions as function of $\Delta M=M(p\bar{p})-2M_p$. left: $\psi(2S)\rightarrow\gamma p\bar{p}$; right: $J/\psi\rightarrow \gamma p\bar{p}$.}
\end{center}
\end{figure}

BES III has recently confirmed~\cite{c18} the existence of a resonance decaying into $\pi^+\pi^-\eta'$ with $M=1836.5^{+6.4}_{-3.7}~\mathrm{MeV}$ and $\Gamma=190\pm39~\mathrm{MeV}$. 
Such a wide resonance could very well decay into $p\bar{p}$ above threshold, and account for the observed threshold enhancement. BES II and we had earlier proposed this possibility, but BES III makes no comment about it in their latest paper~\cite{c18}.

\section{Decays of $\chi_{cJ}$ to $p\bar{p}+\pi^0,\eta$ and $\omega$~\cite{c19}}

The $\chi_{cJ}$ states are strongly populated by E1 radiative decays from $\psi(2S)$. CLEO has recently made measurements of $\chi_{cJ}$ decays to $p\bar{p}+\pi^0,\eta,\omega$~\cite{c19}. The results are presented in Table~\ref{t3}. The errors in these results are factor $>2$ smaller than in the previous measurements. \\

\begin{table}[!tb]
\begin{center}
\begin{tabular}{|l|l|l|l|l|l|l|} \hline\hline
$\mathcal{B}_\chi\times10^4$  &    \multicolumn{2}{|c|}{$\chi_0$}    &    \multicolumn{2}{|c|}{$\chi_1$}     &    \multicolumn{2}{|c|}{$\chi_2$}\\ \hline\hline
                  &   CLEO        &  PDG~\cite{c2}     &   CLEO        &  PDG~\cite{c2}  &   CLEO        &  PDG~\cite{c2}     \\\hline
$\mathcal{B}(\chi_J\rightarrow p\bar{p}\pi^0)$ & $7.8\pm0.7$ &$5.7\pm1.2$   &$1.8\pm0.2$   &$1.2\pm0.5$   &$4.8\pm0.5$   &$4.7\pm1.0$\\\hline
$\mathcal{B}(\chi_J\rightarrow p\bar{p}\eta)$ & $3.7\pm0.5$ &$3.7\pm1.1$   &$1.6\pm0.3$   &$<1.6$   &$1.8\pm0.3$   &$2.0\pm0.8$\\\hline
$\mathcal{B}(\chi_J\rightarrow p\bar{p}\omega)$ & $5.6\pm0.7$ &   &$2.3\pm0.4$   &   &$3.7\pm0.5$   &\\\hline\hline
\end{tabular}
\end{center}
\caption{Branching fractions determined for $\chi_J$ decays to $p\bar{p}\pi^0$, $p\bar{p}\eta$, and $p\bar{p}\omega$.}
\label{t3}
\end{table}

Both sets of measurements, $\psi(2S)\rightarrow p\bar{p}+\gamma,\pi^0,\eta$ in Sec. 5 and $\chi_{cJ}\rightarrow p\bar{p}+\pi^0,\eta,\omega$ in Sec.~6,
are potentially of great value to the future $p\bar{p}$ experimentation at PANDA(GSI).

\section{Multipole Admixtures in Dipole Transitions~\cite{c20}}

If the radiative transitions $\chi_{c1},\chi_{c2}\rightarrow\gamma J/\psi$ are attributed to a single quark, the E1 transitions can have small M2 components, with $a_2=M2/\sqrt{E_1^2+M_2^2}$. Simple predictions are that $a_2(\chi_1)=-(E_\gamma/4m_c)(1+\kappa_c)$, and $a_2(\chi_2)=(-3/\sqrt{5})(E_\gamma/4m_c)(1+\kappa_c)$, where $\kappa_c$ is the anomalous magnetic moment of the charm quark.\\

Previous attempts at SLAC and Fermilab E760/E835 to measure $a_2(\chi_1,\chi_2)$ were limited mainly by statistics, and had large errors.\\

CLEO has recently made a high statistics measurement~\cite{c20}, with the results
\begin{center}
$a_2(\chi_{c1})=(-6.26\pm0.67)\times10^{-2}$, and $~~a_2(\chi_{c2})=(-9.3\pm1.6)\times10^{-2}$. 
\end{center}

The ratio, $a_2(\chi_{c2})/a_2(\chi_{c1})=1.49\pm0.30$ is consistent with $3/\sqrt{5}=1.34$, justifying the hypothesis of a \textbf{single quark transition}. \\

For assumed $m_c=1.5~\mathrm{GeV}$, we get
\begin{center}
$\chi_{c1}:(1+\kappa_c)=0.88\pm0.20,\quad\chi_{c2}:(1+\kappa_c)=1.10\pm0.19$.
\end{center}
Both are consistent with \textbf{the anomalous magnetic moment of the charm quark, $\kappa_c=0$.} \\

In a quenched lattice calculation the Jlab group predicts $a_2(\chi_{c1})=(-20\pm6)\times10^{-2}$, $a_2(\chi_{c2})=(-39\pm7)\times10^{-2}$, factors 3 to 4 larger than our measured values~\cite{c21}.

\section{Interference in Strong and Electromagnetic Decays of $\psi(2S)$ to Pseudoscalar Pairs, $PP=\pi^+\pi^-,K^+K^-$ and $K_S K_L$~\cite{c25}}

Interest in final state interaction (FSI) phases originally arose from CP violation in K decays and B decays. However, it was discovered that large FSI phases are perhaps a general feature. Suzuki~\cite{c22} and Rosner~\cite{c23} have analyzed $J/\psi$ decays into pseudoscalar-vector (PV) pairs, and pseudoscalar-pseudoscalar (PP) pairs, and find that the phase differences between strong and EM decay amplitudes in both PV and PP decays of $J/\psi$, measured as the interior angle $\delta$ of the triangle representing the amplitudes, is large\\

\parbox{4in}
{
$\delta(J/\psi,\psi(2S))_{PP}=\mathrm{cos}^{-1}(\frac{B(K^+K^-)-B(K_SK_L)-\rho B(\pi^+\pi^-)}{2\sqrt{B(K_SK_L)\times\rho\times B(\pi^+\pi^-)}}),$
\\
~~\\ where $\rho=$ phase space factor 
} \ \
\parbox{2in}{
\vspace*{-8pt}
\includegraphics[width=2in]{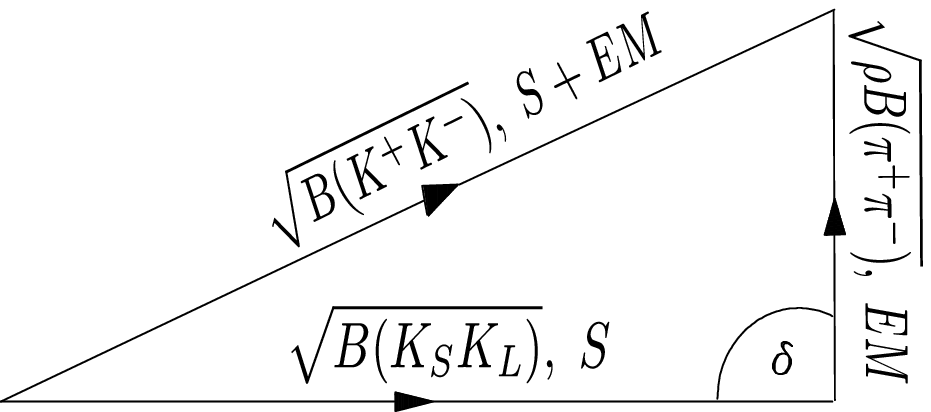}}\\

$$\delta(J/\psi)_{PP}=89.6^\circ\pm9.9^\circ\mathrm{(Suzuki)},~89^\circ\pm10^\circ\mathrm{(Rosner)},~82^\circ\pm9^\circ\mathrm{(PDG2010)}$$

Suzuki~\cite{c24} raised the natural question if the $\sim\pi/2$ phase difference would also be found in the PP decays of $\psi(2S)$. If not, he wondered if it could perhaps explain the so called $\rho\pi$ (PV) problem: $\mathcal{B}(\psi(2S)\rightarrow\rho^0\pi^0)/\mathcal{B}(J/\psi\rightarrow\rho^0\pi^0)\approx 0.6\%$, instead of the pQCD expected value of $\sim13\%$.\\

Previous measurements with small statistics $\psi(2S)$ data indicated large phase difference, $\delta(\psi(2S))_{PP}$, but with large errors, mainly due to the very small $\mathcal{B}(\psi(2S)\rightarrow\pi^+\pi^-)$, whose strong decay is forbidden by isospin conservation. \\

\begin{figure}
\begin{center}
\includegraphics[width=1.8in]{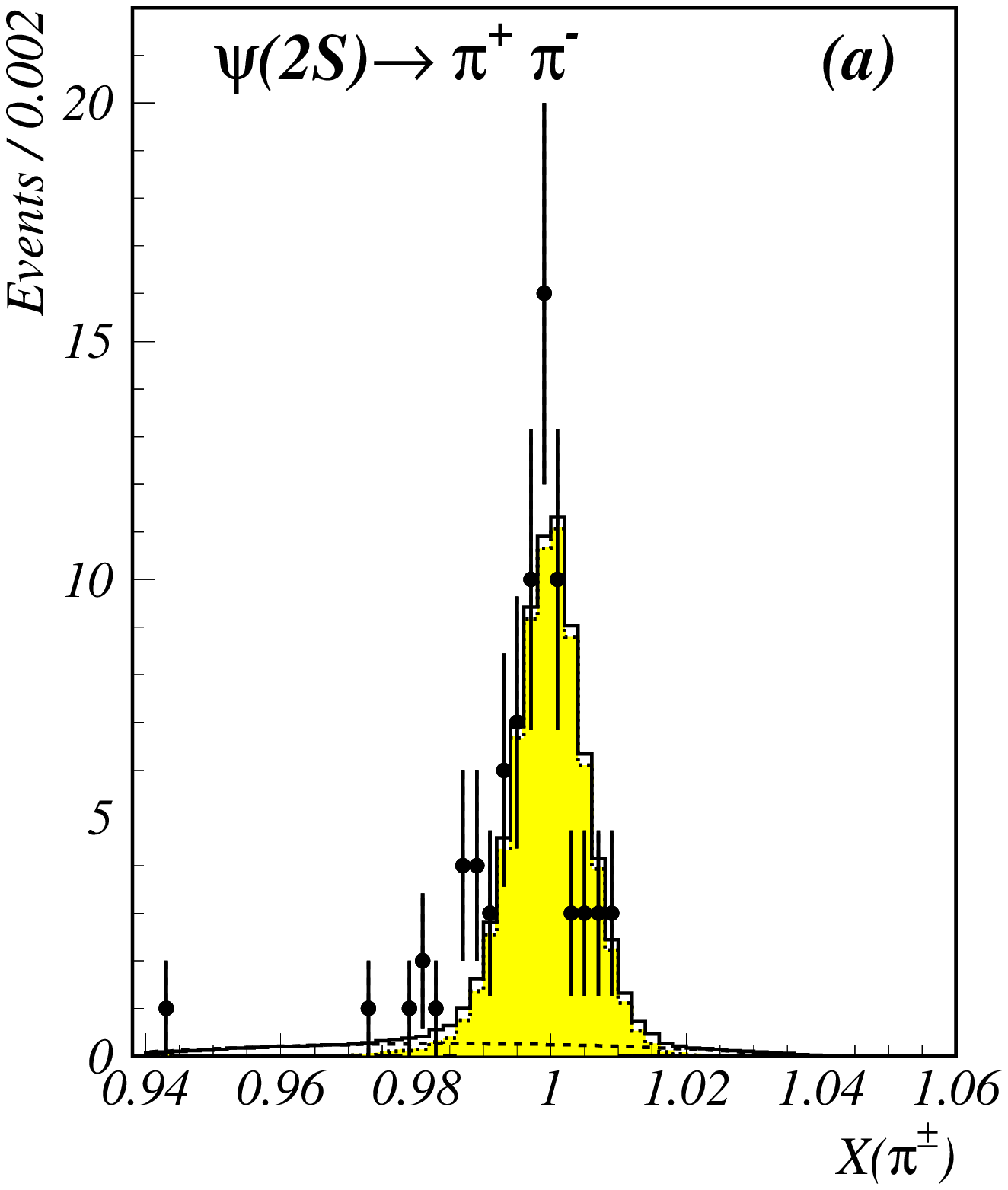}
~~~
\includegraphics[width=1.8in]{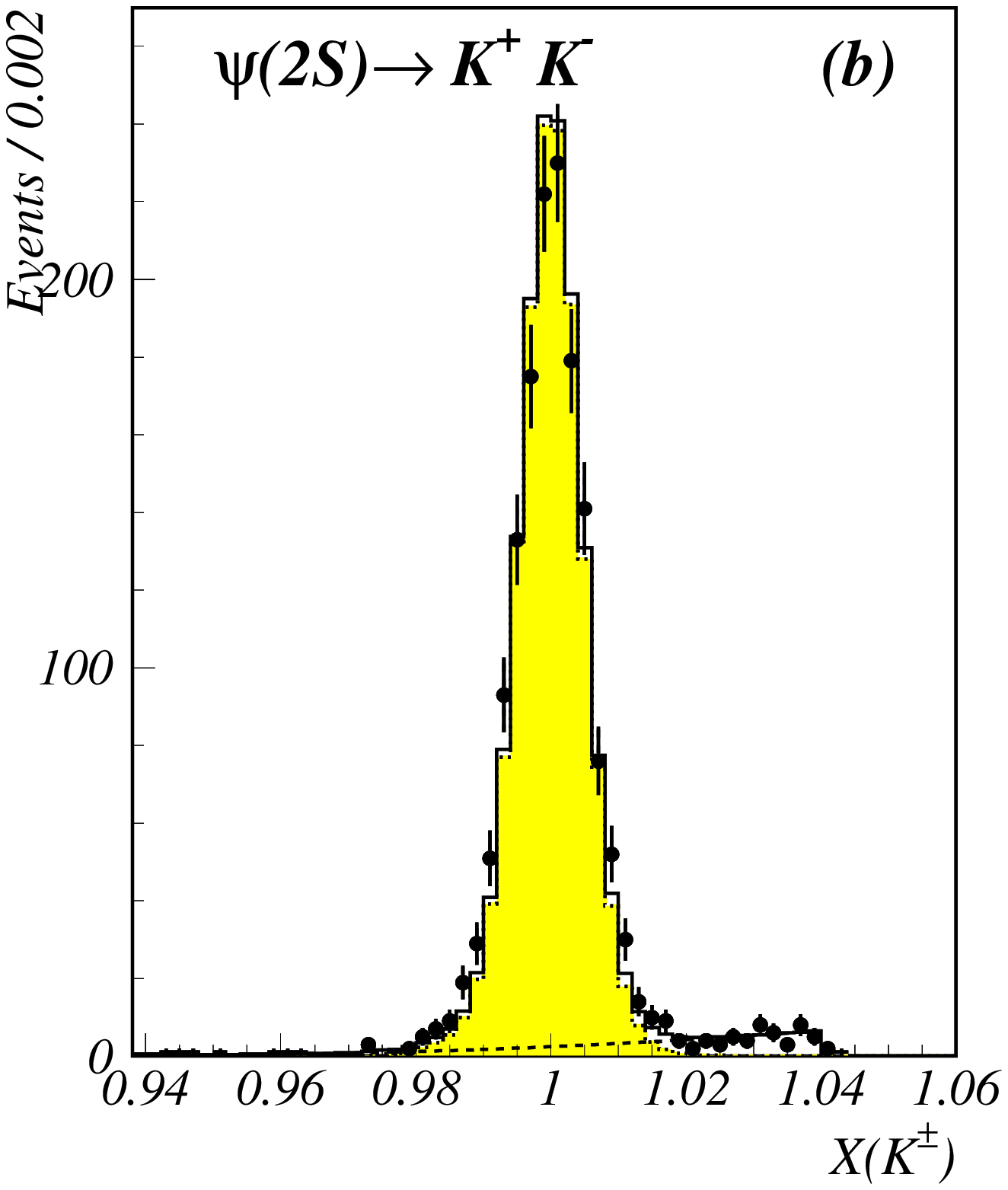}
~~~
\includegraphics[width=1.8in]{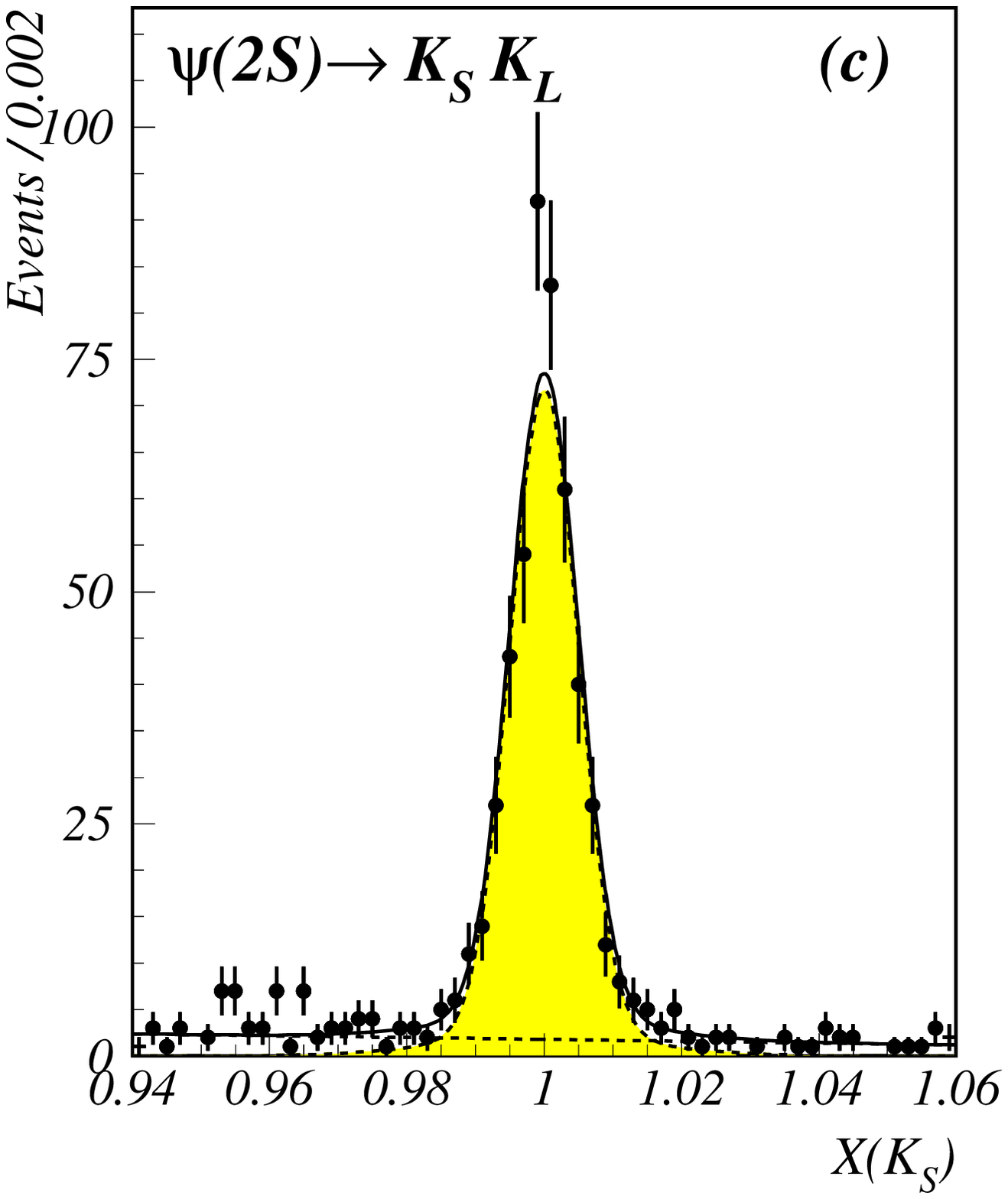}
\end{center}
\caption{Event distributions as functions of $X(h)\equiv E(h)/E(\mathrm{beam})$ for $h=\pi^\pm,K^\pm$, and $K_s$.}
\end{figure}

CLEO has now made a new measurement with 24.5 million $\psi(2S)$, and I present the preliminary results here~\cite{c25}. The event distributions obtained for the three decays $\psi(2S)\rightarrow\pi^+\pi^-,K^+K^-$ and $K_SK_L$ are shown in Fig. 6. The preliminary results for the measured branching fractions for the decays are listed in Table~\ref{t4}. These lead to a more precise result, $\delta(\psi(2S))_{PP}=114^\circ\pm11^\circ$.\\

\begin{table}[!tb]
\begin{center}
\begin{tabular}{|c|c|c|c|c|} \hline\hline
                  &   DASP       & BES            &   CLEO  &  This \\
                  &   1979       & 2004           &   2005  &  analysis   \\\hline\hline
$\mathcal{B}(\pi^+\pi^-)\times10^5$  & $8\pm5$    &$0.84\pm0.65$   &$0.8\pm0.8$   & $0.72\pm0.24$  \\ \hline
$\mathcal{B}(K^+K^-)\times10^5$     & $10\pm7$  &$6.1\pm2.1$   &$6.3\pm0.7$   &$7.49\pm0.43$ \\\hline
$\mathcal{B}(K_SK_L)\times10^5$     & --  &$5.24\pm0.67$   &$5.8\pm0.9$   & $5.31\pm0.43$ \\\hline
$\delta(\psi(2S))_{PP}$ & --   &$(91\pm35)^\circ$$^*$   &$(87\pm20)^\circ$$^*$   & $(114\pm11)^\circ$ \\\hline\hline
\end{tabular}\\
\vspace*{5pt}
$^*$ Recalculated $\qquad\qquad\qquad\qquad\qquad\qquad\qquad\qquad\qquad\qquad\qquad\qquad$
\end{center}
\vspace*{-10pt}
\caption{Branching fractions determined for $\psi(2S)$ decays into pseudoscalar pairs, $\pi^+\pi^-,K^+K^-$ and $K_SK_L$, and the resulting interference angle $\delta(\psi(2S))_{pp}$.}
\label{t4}
\end{table}

In summary, both $J/\psi$ and $\psi(2S)$ decays to pseudoscalar pairs give large phase difference between strong and EM amplitudes. The difference between $\delta(J/\psi)=82^\circ\pm9^\circ$ and $\delta(\psi(2S))=114^\circ\pm11^\circ$ is $2.3\sigma$. Question: Is this significant?

\section{Summary}

We have reported new results from the analysis of CLEO data for
$\psi(2S)$, $\psi(4170)$, $\Upsilon(2S)$, and $\Upsilon(3S)$. These include:
\begin{enumerate}
\item Branching fractions for $\psi(2S)\to\pi^0 h_c(^1P_1)$.
\item Production of $h_c(^1P_1)$ in $e^+e^-(4170) \to \pi^+\pi^- h_c(^1P_1)$.
\item Branching fractions for $\Upsilon(3S)\to \gamma \chi_{b1.b2}(1P)$.
\item Decays of $\psi(2S)$ and $J/\psi\to p\bar{p}+\gamma$, $\pi^0$,
and $\eta$, and search for $p\bar{p}$ threshold enhancements.
\item Multipole admixtures in $\psi(2S)\rightarrow\gamma\chi_J,\chi_J\rightarrow\gamma J/\psi$ dipole transitions.
\item Interference between strong and electromagnetic amplitudes in
$\psi(2S)$ decays to pseudoscalar pairs, $\pi^+\pi^-$, $K^+K^-$ and $K_SK_L$.
\end{enumerate}
These results pose several interesting physics questions.  Among these are:
\begin{itemize}
\item Why $\Delta M_{\mathrm{hf}}(1P)\equiv\langle M(^3P_J) \rangle -M(^1P_1) =0$, if $\langle M(^3P_J) \rangle \neq M(^3P)$?
\item What hadronic decays account for $\mathcal{B}(h_c\to\mathrm{hadrons})\approx50\%$ ?
\item Why is the $p\bar{p}$ threshold enhancement seen in $J/\psi$ decay not seen in $\psi(2S)$ decay?
\item What is the significance of the $2.3\sigma$ difference seen in the interference angle between strong and electromagnetic PP decays of $J/\psi$ and $\psi(2S)$.
\end{itemize}

\newpage


%

}  



\begin{thebibliography}{99}
  

\bibitem{c1} A. Tomaradze, Proc. Hadron 2009, AIP Conf. Proc. 1257, 197 (2010).
\bibitem{c2} PDG 2010: Particle Data Group, ``Review of Particle Physics", Journal of Physics G, 37, 075021 (2010).
\bibitem{c3} S. K. Choi et al. (Belle Collaboration), PRL 89, 102001 (2002).
\bibitem{c4} D. M. Asner et al. (CLEO Collaboration), PRL 92, 142001 (2004).
\bibitem{c5} B. Aubert et al. (BaBar Collaboration), PRL 92, 142002 (2004).
\bibitem{c6} J. L. Rosner et al. (CLEO Collaboration), PRL 95, 102003 (2005).
\bibitem{c7} B. Aubert et al. (BaBar Collaboration), PRL 101, 071801 (2008).
\bibitem{c8} G. Bonvicini et al. (CLEO Collaboration), PRD 81, 031104(R) (2010).
\bibitem{c9} S. Dobbs et al. (CLEO Collaboration), PRL 101, 182003 (2008).
\bibitem{c10} M. Ablikim et al. (BES III Collaboration), PRL 104, 132002 (2010).
\bibitem{c11} J. Y. Ge et al. (CLEO Collaboration), arXiv: 1106.3558 [hep-ex].
\bibitem{c12} G. S. Adams et al. (CLEO Collaboration), PRD 80, 051106(R) (2009).
\bibitem{c13} T. Pedlar et al. (CLEO Collaboration), arXiv: 1104.2025 [hep-ex], accepted by PRL.
\bibitem{c14} I. Adachi at al. (Belle Collaboration), arXiv: 1103.3419 [hep-ex].
\bibitem{c15} D. M. Asner et al. (CLEO Collaboration), PRD 78, 091103(R) (2008).
\bibitem{c16} M. Kornicer et al. (CLEO Collaboration), PRD 83, 054003 (2011).
\bibitem{c17} J. P. Alexander et al. (CLEO Collaboration), PRD 82, 092002 (2010).
\bibitem{c18} M. Ablikim et al. (BES III Collaboration), PRL 106, 072002 (2011).
\bibitem{c19} P. U. E. Onyisi et al. (CLEO Collaboration), PRD 82, 011103(R) (2010).
\bibitem{c20} M. Artuso et al. (CLEO Collaboration), PRD 80, 112003 (2009).
\bibitem{c21} J. J. Dudek et al., PRD 79, 094505 (2009).
\bibitem{c22} M. Suzuki, PRD 60, 051501(R) (1999).
\bibitem{c23} J. L. Rosner, PRD 60, 074029 (1999).
\bibitem{c24} M. Suzuki, PRD 63, 054021 (2001).
\bibitem{c25} K. K. Seth, for the CLEO Collaboration, this talk (preliminary results).
\end{thebibliography}
\end{document}